\documentclass[sigconf]{acmart}

\usepackage[linesnumbered,ruled,vlined]{algorithm2e}
\usepackage{multicol}
\usepackage{comment}
\usepackage{array}
\usepackage{eqparbox}
\usepackage{color}
\usepackage{subcaption}

\usepackage{caption} 

\input{savespace}
\newcommand{\fullversion}[1]{}

\newcommand{\mycomment}[1]{}
\newcommand{\mysubsection}[1]{\vspace{1ex}\noindent\textbf{#1.}\xspace}

\SetFuncSty{text}
\SetArgSty{text}
\SetKw{OR}{or}
\SetKw{AND}{and}
\SetKw{NOT}{not}
\SetKw{TRUE}{true}
\SetKw{FALSE}{false}
\SetKw{NULL}{nil}
\SetKw{Break}{break}
\SetKw{Continue}{continue}
\SetKwInOut{Input}{Input}
\SetKwInOut{Output}{Output}
\SetKwFor{RepTimes}{repeat}{times}{end}

\AtBeginDocument{%
  \providecommand\BibTeX{{%
    \normalfont B\kern-0.5em{\scshape i\kern-0.25em b}\kern-0.8em\TeX}}}

\copyrightyear{2022}
\acmYear{2022}
\setcopyright{acmcopyright}

\acmConference[SIGIR '22] {Proceedings of the 45th International ACM SIGIR Conference on Research and Development in Information Retrieval}{July 11--15, 2022}{Madrid, Spain.}
\acmBooktitle{Proceedings of the 45th International ACM SIGIR Conference on Research and Development in Information Retrieval (SIGIR '22), July 11--15, 2022, Madrid, Spain}
\acmPrice{15.00}
\acmISBN{978-1-4503-8732-3/22/07}
\acmDOI{10.1145/3477495.3531678}
\begin{document}
\fancyhead{}
\title{Table Enrichment System for Machine Learning}


\author{Yuyang Dong}
\affiliation{%
  \institution{NEC Corporation, Japan}
  \country{}
}
\email{dongyuyang@nec.com}

\author{Masafumi Oyamada}
\affiliation{%
  \institution{NEC Corporation, Japan}
  \country{}
}
\email{oyamada@nec.com}


\begin{abstract}
Data scientists are constantly facing the problem of how to improve prediction accuracy with insufficient tabular data.
We propose a table enrichment system that enriches a query table by adding external attributes (columns) from data lakes
and improves the accuracy of machine learning predictive models.
Our system has four stages, join row search, task-related table selection, row and column alignment, and feature selection and evaluation,
to efficiently create an enriched table for a given query table and a specified machine learning task.
We demonstrate our system with a web UI to show the use cases of table enrichment.

\end{abstract}



\maketitle

\section{Introduction}

Given a table and ML prediction task,
many recent works on Auto-ML \cite{automl-survey} are designed to automatically select the best features and models.
However, users still have trouble if their tables do not contain enough signals (features) for training an ML model with satisfying accuracy.
Features are always lacking in practical data analysis scenarios,
for example, when a data analyst plans to classify the category of products but only has a few basic attributes such as the product name.
The description of products may help to better predict global sales,
and these important features may sleep in data lakes such as web scraping tables on E-commerce sites. 

With the trends of open data and the growth of data lake solutions,
we can easily access and obtain 
a huge amount of tabular data from various domains in data lakes 
(e.g., the WDC Web Table Corpus \cite{URL:WDC}).
This naturally comes out with a research problem:
\textbf{Can we build a system that automatically enriches a query table with external columns from data lakes
and improve the ML prediction task?}

\mysubsection{Related works}
We need three steps to achieve table enrichment: table retrieval, table join, and ML evaluation.
There are many related works on table retrieval \cite{webtable-ERA-survey, webtables} 
with different end tasks such as column and row extension \cite{find-related-table, msj, entitable, concept-expansion}, 
schema matching \cite{table-union-search, table-stitch},
table filling \cite{infogather,infogather+}, and knowledge graph building \cite{tablenet, web-table-interpretation}. 

Regarding table retrieval and table join, because the goal of the enrichment is to connect 
a query table to the external tables in data lakes by joining them with record value matching.
The most basic requirement is to retrieve joinable tables from data lakes \cite{josie, pexeso}
There is also other research to retrieve related tables but not join. 
Zhang et al. studied the problem of finding related tables in data lakes for interactive data science \cite{inter-ds-sigmod}
by a composed criteria with a join rate, new column rate, and null value decrement.
Mahdi et al. \cite{cocoa} proposed a COCOA system to efficiently compute Spearman's correlation coefficient after the join tables are mapped.
Aécio et al. \cite{join-corr} studied the problem of joining-correlation query,
which can efficiently find tables that can be joined and also contains 
correlated values with a query table.
Nevertheless, the above works focus on retrieving tables, 
and they do not cover the following steps of enriching a table to improve the machine learning accuracy.

Regarding ML evaluation, some works target the problem of feature selection 
in which dozens of candidate joinable tables are given.
Kumar et al. \cite{to-join-not-join} proposed a TR-rule based on the bound of the VC dimension \cite{URL:vc}
to skip unnecessary table joins. 
However, the ML task and model are limited since the theory of the VC dimension is only for classification.
Chepurko et al. \cite{arda} studied efficient feature selection over candidate tables with sampling techniques and proposed a feature selection method that 
ranks features with a random injection of noise. 
The above works assume a few candidate tables are already given and process them one by one.
Hence, they do not consider how to retrieve useful tables from data lakes,
so it is not efficient to use them to process a large number of tables.

In conclusion, there is no generalized table enrichment system,
and this paper fills this gap by proposing an end-to-end system 
that covers the whole pipeline with table discovery, table augmentation, and ML model evaluation.

\begin{figure*}[t]
  \centering
  \includegraphics[width=0.8\linewidth]{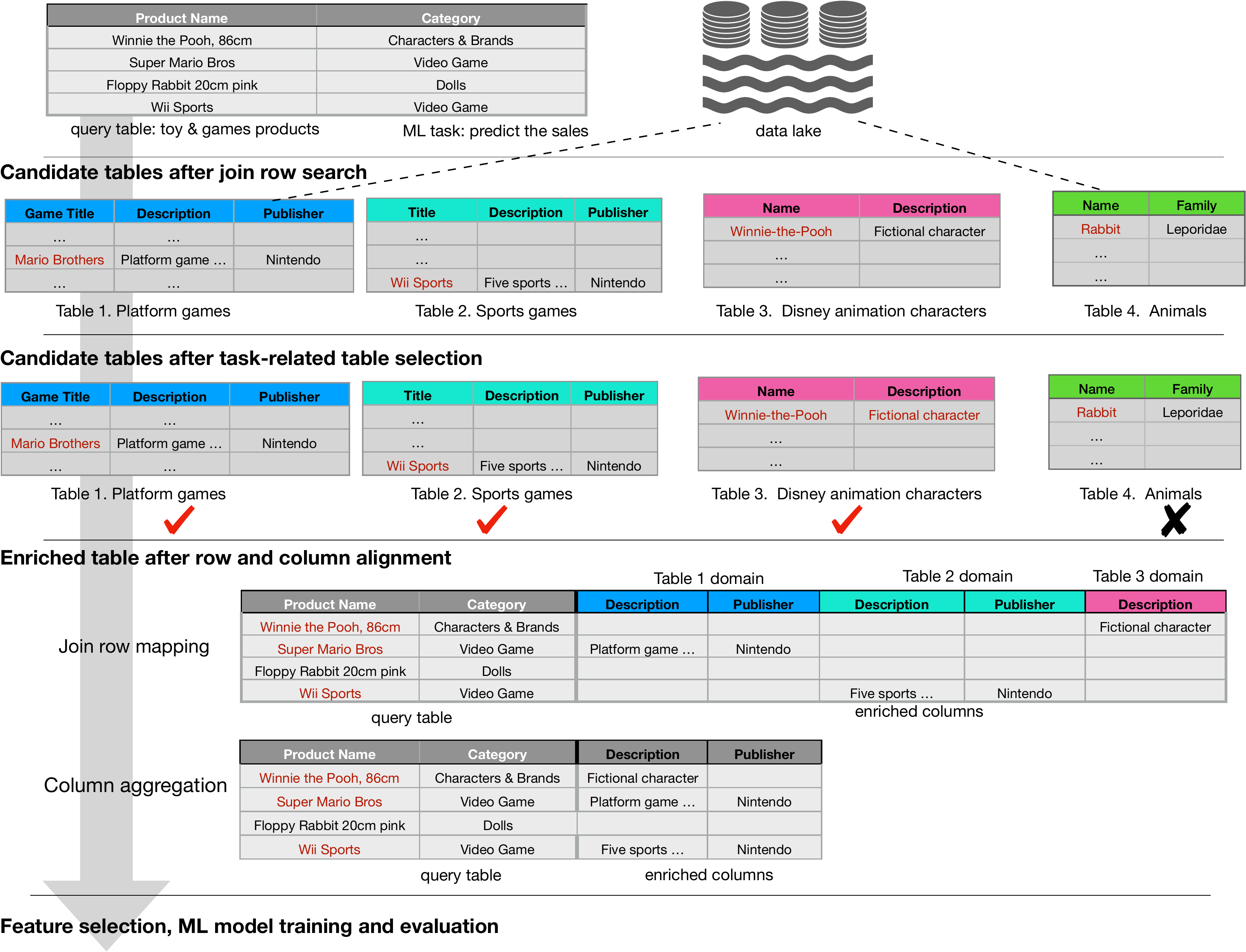}
  \caption{Example of table enrichment for ML.}
  \label{fig:overall-example}
\end{figure*}


\mysubsection{Problem statement}
\label{sec:prob:state}
Table enrichment system takes as input the user's query table $q = (q.c, q.y,$ $q.column\_names, q.data)$,
where $q.data$ is the whole tabular data and $q.column\_names$ is a string list of column names.
$q.y$ is a target column for prediction, and $q.c$ is a query column(s) for join.
$T = \{t_1, t_2, ..., t_n\}$ is a collection of target tables.
Besides the target table data itself, we assume a target table $t$ also has two types of metadata, 
the table title and table context, which can be easily retrieved from the source of tables.
Therefore,  $t = (t.title, t.context, t.column\_names, t.data)$.

\textbf{Table Enrichment Problem}: Given a query table $q$,
the goal of {\em table enrichment} is to create an enriched table $q'$
using the data from a collection of target tables $T$.
Without loss of generality, we keep the query table unchanged and fix the number of rows, 
and we only add columns (features) to it.  
The enriched table $q'$ will improve the accuracy of predicting $q.y$ compared to only predicting with $q$, that is,
$Acc(q',q.y) > Acc(q, q.y)$.

\section{Table Enrichment System}
Figure \ref{fig:overall-example} gives a running example of our table enrichment system.
There is a query table of toy \& game products, and the task is to classify the categories of products
First, we retrieve four candidate tables that can join with the query table.
Then, Tables 1,2,3 are selected since the table they are related to toy \& games,
while Table 4 ``animal" is an irrelevant one and we filter it.
We also aggregate the values in the same column name to make a compact and clean enriched table.
Finally, we input this enriched table into the ML evaluation phase with feature selection, model training, and evaluation.

\subsection{Join row search}

Given a query column $q.c$ and a collection of target tables $T$, we retrieve candidate join rows to $q.c$ from the tables in $T$.
For each cell in $q.c$, we retrieve similar cells in $T$, and the rows of these cells are candidate join rows.

For a query cell value $a$ in $q.c$ and a cell value $b$ in $T$, 
even if theirs rows refer to the same entity, the cell values may have different presentation styles,
such as ``Mario Bros" and ``Super Mario Brothers."
Therefore, we leverage Jaccard, BM25, and semantic similarities to cover different levels of fuzzy matching between query cells and target cells.
Jaccard similarity treats $a$ and $b$ as sets of words,
and we use it to retrieve join rows that share the same words with a query.
\begin{equation}
\label{equ:jac}
    Jaccard(a, b) = \frac{|a \cap b|}{|a \cup b|}
\end{equation}

BM25 similarity is an optimization of TF-IDF that treats $a$ and $b$ as documents,
and we use it to retrieve join rows on the basis of the word frequency and document frequency.
\begin{equation}
\label{equ:jac}
    BM25(a,b) = \sum_{i=1}^{n} IDF(a_i) \cdot \frac{TF(a_i, b)\cdot (m + 1)}{TF(a_i, b) + m(1-h + h \cdot \frac{|b|}{Avgdl})}
\end{equation}
where $TF(a_i, b)$ is the term frequency in $b$ of the term $a_i \in a$,
and $IDF(a_i)$ is the inverse document frequency of the term $a_i$.
$m$ and $h$ are constants,
and $avgdl$ is the average document length in the text collection from which documents are drawn.

Semantic similarity involves using a pre-trained language model such as fastText \cite{fasttext-paper} and BERT \cite{bert-paper}
to encode the string values of $a$ and $b$ into vector representations,
and we compute the Euclidean distance between vectors as similarities.
It is used to retrieve join rows that may not be similar in a string but are semantically similar.
\begin{equation}
\label{equ: semantic}
    Semantic(a,b) = ||encode(a) - encode(b)||_2
\end{equation}

We first search top-$k$ similar cells (rows) from target tables $T$ with Jaccard, BM25, and semantic similarity independently.
Then, we take the union of all candidate rows as the final output.
To efficiently index and retrieve join rows from large tables,
we implement the join row search engine with Elasticsearch \cite{URL:elasticsearch}.
Elasticsearch is a powerful search engine that supports BM25 with Lucene \cite{URL:lucene} implementation,
Jaccard similarity with Minhash token and LSH implementation,
and kNN search with nmslib \cite{nmslib} implementation.
Therefore, a similarity search can be processed very fast with these sophisticated indices.
Another reason we used 
Elasticsearch is because it is a disk-based distributed engine,
and it is easy to increase the capacity without extra maintenance costs.

\subsection{Task-related table selection}
\label{sec:tab:selection}

There are large numbers of candidate tables after the join row search.
It is very time-consuming to create a large enriched table with them.
More seriously, the candidate tables also contain many irrelevant tables that
they can join with the query table, but their information cannot contribute to the ML task.
Adding these noisy data (columns) will weaken the performance of feature selection algorithms \cite{fs-survey} 
and decrease the accuracy of an ML model.
We propose a further selection to pick up task-related tables.
\begin{figure}[t]
  \centering
  \includegraphics[width=\linewidth]{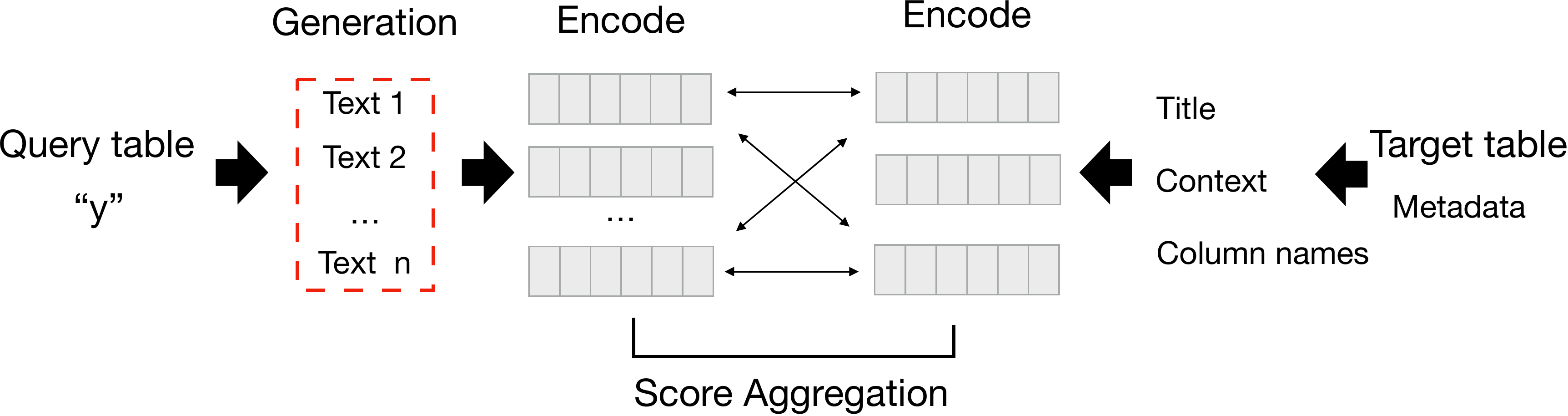}
  \caption{Text-based query and semantic matching.}
  \label{fig:select}
\end{figure}

We formalize this task-related table selection problem as a top-$m$ table retrieval problem with text-based query and semantic matching \cite{ad-hoc-table}. 
The purpose of using a text-based query is to match a related table in accordance with
the description of the ML task.
The goal of semantic matching is to go beyond lexical matching by using pre-trained language models to encode both query text and
target text as vectors into a semantic space.

For a query table,  we allow users to input the description text of their ML task as query text.
We also propose generating multiple query texts if task description text is not available.
The key point is to input the target information (e.g., the ``category" of products) into the query text.
We use the column names concatenation as query text.
The names of columns contain both attribute and target information, 
e.g., ``product name, product category".
For a target table, we set the attributes $t.title$, $t.context$, and $t.column\_names$ as target texts.
To evaluate the related score of a target table to the query table,
we take the maximum semantic similarity score between a query text $q_{text}$ 
and all target texts.
Figure \ref{fig:select} gives an image of the semantic matching of a query table with a target table.
There are many ways to aggregate the pairwise scores between encoded vectors, and we take a maximum 
aggregate function in our experiment.
\begin{equation}
    score_{max}(q,t) = max(score(q_{text}, t_{text[i]}))
\end{equation}

We can control the selected table number to retrieve the top-$m$ related tables after join row search.
This helps us to narrow down the number of candidate tables to hundreds of tables.

\subsection{Row and column alignment}
\label{sec:row:column:alignment}

The next step is to map these rows and join them to the query table.
We join the candidate rows and align the rows if they are from the same table.
Note that for each query row, multiple join rows may be retrieved from the same target table.
In this case, we map the row with the highest similarity score since a target table usually has 
only one matching row (entity) to a query row.

As shown in Figure \ref{fig:overall-example},
the enriched table may contain a large number of columns with many empty values.
This is because different query rows may not join the same target table, 
and the row number of a target table may also differ from the query table.
The size of enriched columns determines the processing time,
and empty values will decrease the performance of the feature selection algorithm and predictive models.

To achieve high performance with a short processing time,
we propose to aggregate columns to decrease the column numbers and impute most of the empty values.
We use two methods:
(a) hard aggregation: we aggregate the columns within the exact same column name;
(b) soft aggregation: we first encode the column names into vectors with pre-trained language models such as fastText\cite{fasttext-paper} and BERT \cite{bert-paper}. Then, we cluster the columns through the k-means algorithm on the encoded vectors.
Last, we aggregate the columns in the same cluster.
Having more compact enriched columns does not always result in better accuracy in the final ML task, so
the cluster number should be tuned in practice.
During aggregation, the values may conflict at the same row position, 
and we aggregate the conflicting values by string concatenation or averaging numerical values. 
Figure \ref{fig:overall-example} also shows the result after column aggregation.
We can see that it is much cleaner with fewer empty values than
the result of join row mapping.

\begin{figure*}
  \centering
  \includegraphics[width=0.85\linewidth]{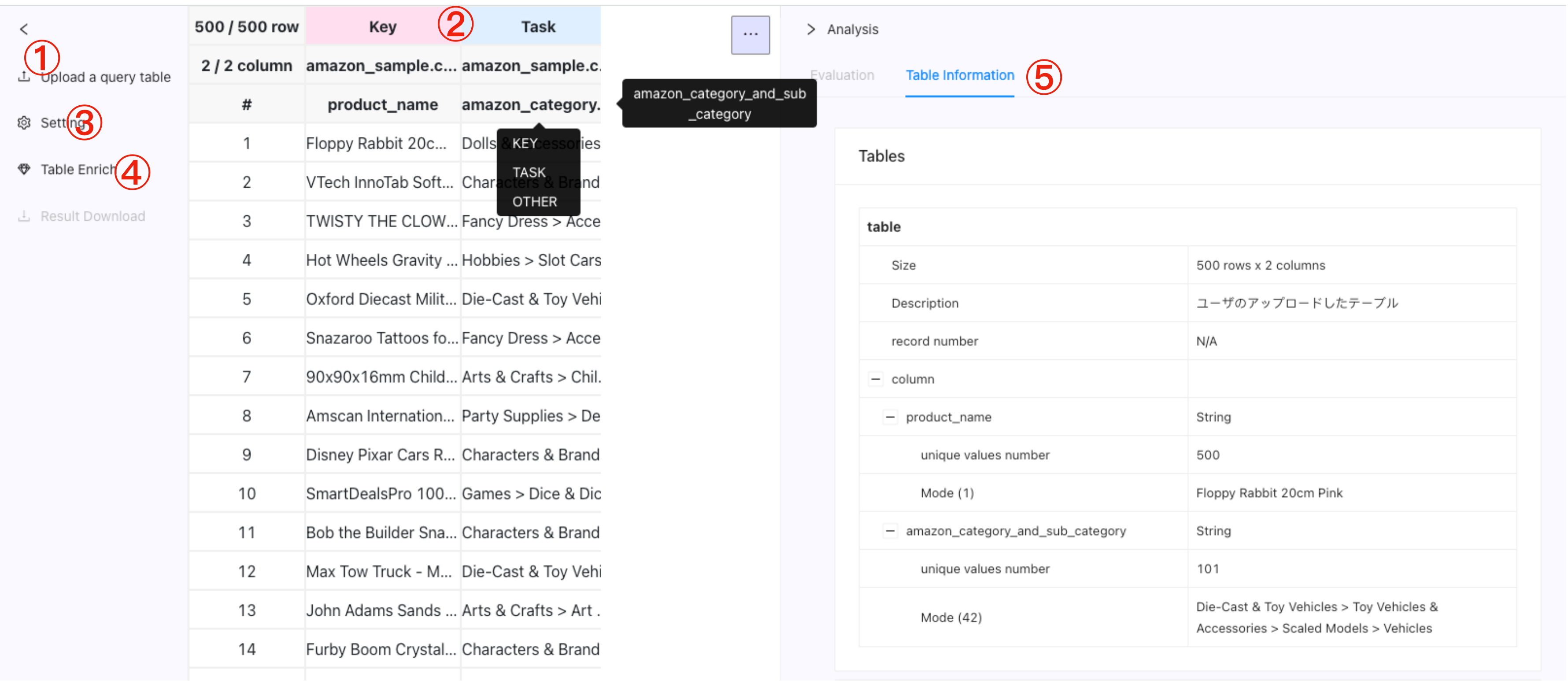}
  \includegraphics[width=0.85\linewidth]{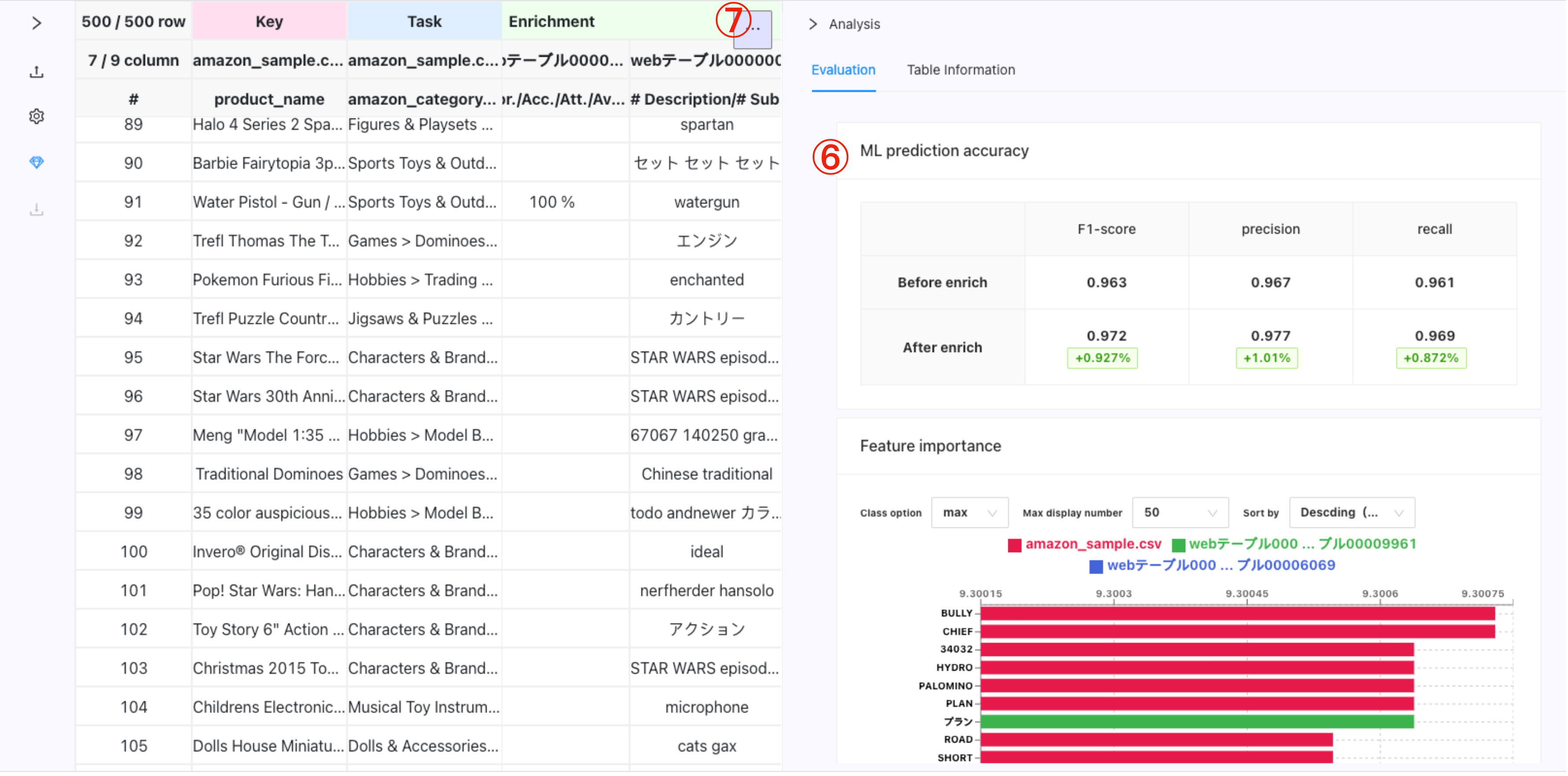}
   \caption{Web UI of table enrichment demo. Steps of our guided demonstration are marked in red circle.}
  \label{fig:demo}
\end{figure*}

\subsection{Feature selection}
Finally, we process feature selection to filter noisy columns.
Our system offers several feature selection methods from the scikit-learn \cite{URL:skleanfs}
including using F-value, forward selection, backward selection, recursive
feature elimination, the random forest model, and linear model with L1 regularization.
We let users tune and select the best one for their data.

\section{Demonstration}

We will demonstrate our table enrichment system on a real-world dataset.
There are two query tables, the AMAZON toy dataset \cite{URL:amazon} and the Car Sales dataset \cite{URL:carsale},
and we enrich them with a data lake that contains 10M WDC web tables \cite{URL:WDC}.
Each step is marked in a red circle in Figure \ref{fig:demo}.
The following description is for the AMAZON toy dataset,
and a video of both two datasets with our demonstration can be found here 
\footnote{https://youtu.be/HXikNjblUwU}.

\mysubsection{Step 1. Upload query table}
First, the user uploads the query table (train and test tables) in the CSV file format. 
The uploaded table will be displayed and statistical information such as row and column numbers can be seen.
In our guided demonstration, we upload the AMAZON toy dataset \cite{URL:amazon}.

\mysubsection{Step 2. Specify the key column and the task column}
The user specifies a column as the key column that
is used as the join key to connect external tables.
Then the user also specifies another column as the task column 
which contains the values to predict with the ML algorithm.
In our guided demonstration, we specify the ``product name" column as the key column and the ``category" for the task column.

\mysubsection{Step 3. Table enrichment parameter setting}
Next, the user sets the parameters for table enrichment, including the domain source of the target table,
similarity function for join row searching,
the methods for alignment column and feature selection.
In our guided demonstration, we set the target table dataset as WDC web tables \cite{URL:WDC}, and use the default parameter setting.

\mysubsection{Step 4. Run table enrichment and get the enriched table}
Then, the user clicks the ``Table Enrich" button to run the table enrichment and sees the results.
In our guided demonstration, 
the uploaded AMAZON table with enriched columns of product descriptions.

\mysubsection{Step 5. Information of retrieved target tables}
After table enrichment, the user can see where those enriched columns come from.
The information and metadata of the retrieved target table and metadata can be shown.
In our guided demonstration, we show the metadata of the title and context of the retrieved target tables.
Our metadata also contains the source URL.

\mysubsection{Step 6. ML performance improvement and feature importance}
Then, the user can see the ML prediction accuracy result before (using only the query table itself) and after table enrichment.
The user can also confirm the effectiveness of table enrichment by checking the bar chart 
showing the feature importance under the ML score results.
In our guided demonstration, we show the classification evaluation of product category classification with precision, recall, and F1-score.
The feature importance is also ranked with a bar chart with different colors to show that
the enriched features also contribute to the ML performance.

\mysubsection{Step 7. Insights for ML performance}
Last, the user can see the record level prediction results with the display filter in our system.
The user can observe the difference in prediction results before and after table enrichment.
This option is for the classification task.
In our guided demonstration, we filter the records to show the different classification on AMAZON toy dataset.
After this, the user can adjust the parameters for running the next loop of table enrichment
and achieving a better ML performance.

\section{Conclusion}
We proposed a table enrichment solution to retrieve, select, join, and aggregate candidate tables step by step.
We show the use case of our system with a demo on a web UI.
As future work, we plan to extend our system with existing data clean techniques 
to create more valuable enriched tables for ML.

\section*{Acknowledgement}
We thank Dr. Takuma Nozawa and Masafumi Enomoto (NEC Corporation) for the discussions of this research.

\bibliographystyle{ACM-Reference-Format.bst}
\bibliography{sample-sigconf.bib}

\end{document}